\documentclass{aa}
\usepackage{graphicx}
\usepackage{lscape}
\usepackage{epsfig}
\begin{document}

\title{Towards a pure ZZ Ceti instability strip
{\footnote{Partially based on
observations at Observat\'orio do Pico dos Dias/LNA; the Southern Astrophysical
Research telescope, a collaboration between CNPq-Brazil, NOAO, UNC, and MSU; and
McDonald Observatory of The University of Texas at Austin.}}}

\author{
B. G. Castanheira\inst{1}\inst{,2},
S. O. Kepler\inst{1},
A. F. M. Costa\inst{1},
O. Giovannini\inst{3},
E. L. Robinson\inst{2},
D. E. Winget\inst{2},
S. J. Kleinman\inst{4},
A. Nitta\inst{5},
D. Eisenstein\inst{6},
D. Koester\inst{7},
M. G. Santos\inst{1}
}

\offprints{barbara@if.ufrgs.br}
\institute{
Instituto de F\'{\i}sica, Universidade Federal do Rio Grande do Sul,
  91501-900  Porto-Alegre, RS, Brazil\\
\and Department of Astronomy and McDonald Observatory,
  University of Texas,
  Austin, TX 78712, USA\\
\and Departamento de F\'{\i}sica e Qu\'{\i}mica, Universidade
de Caxias do Sul, Caxias do Sul, RS,  Brazil
\and Subaru Telescope, National Astronomical Observatory of Japan, 650 North 
A'ohoku Place, Hilo, HI 96720, USA \\
\and Gemini Observatory, Northern Operations Center, 670 North A'ohoku Place, 
Hilo, HI 96720, USA \\
\and Steward Observatory, University of Arizona,
933 N. Cherry Ave.  Tucson, AZ 85721, USA\\
\and Institut f\"ur Theoretische Physik und Astrophysik, Universit\"at Kiel,
  24098 Kiel, Germany\\
}

\date{Received 22/06/2006; accepted 15/09/2006}

\abstract
{We have observed again two stars inside the ZZ Ceti instability strip that
were previously classified as not-observed-to-vary (NOV) by Mukadam et al. 
(2004) and found them to be low-amplitude variables. Some evidence points
to a pure ZZ Ceti instability strip; other evidence contests it.}
{The two stars previously classified as NOV have Sloan Digital Sky Survey
(SDSS) spectroscopic effective 
temperatures that place them inside the ZZ Ceti instability strip, and they
were ``contaminating" the strip as constant stars, which could indicate that
the instability strip was no longer a simple evolutionary stage. A pure
instability strip indicates that pulsation is a normal phase which all DAs
must go through.}
{We used effective temperatures derived from SDSS optical spectra by comparing
them with model atmospheres to look for pulsators through time-resolved
photometry and stars previously classified as NOV.}
{Our new results indicate, but do not prove, a pure instability strip, 
because there are still other NOV stars that need to be observed again. 
Additionally, we have discovered five other ZZ Ceti stars based on their
effective temperatures.} 
{}

\keywords{(Stars): white dwarfs, Stars: variables: general, Stars: oscillations}

\titlerunning{Towards a pure ZZ Ceti instability strip}
\authorrunning{B. G. Castanheira  et al.}

\maketitle

\section{Introduction}

The observational DA instability strip (ZZ Cetis) ranges in effective
temperature ($T_{\mathrm{eff}}$) from 12\,270 to 10\,850\,K (Gianninas, 
Bergeron, \& Fontaine 2005; Mukadam et al. 2004). The current best theoretical
models (Fontaine et al. 2001; Bradley 1996; Arras, Townsley, \& Bildsten 2006)
predict that the ZZ Ceti instability strip should be pure: all the DA white
dwarf stars, with temperatures within the instability strip limits should 
pulsate, with a small dependency on mass. The theory agreed extremely well
with observations until the search for new pulsators was extended to white
dwarfs discovered by the Sloan Digital Sky Survey (SDSS), which reaches a 
fainter population (16 $<$ g $<$ 20) than the old sample. Using accurate
temperature determination either from optical or ultraviolet spectra, the 
instability strip using only the old bright sample did not contain any non 
pulsator (e.g. Bergeron et al. 2004; Gianninas, Bergeron, \& Fontaine 2005).

However, in the searches conducted by Mukadam et al. (2004) and Mullally et al.
(2005), a substantial fraction of the DA white dwarf stars with temperatures
placing them inside the instability strip do not show pulsation above their
detection limit, so were called not-observed-to-vary (NOV). 
Many questions were raised based on these new observations. Is the instability
strip pure? Which are the physical mechanisms that can prevent pulsation? What
does this particular sample have that is different from the old one? Can the H
layer be extremely thin so that pulsations will not happen? Can even weak 
magnetic fields prevent pulsation? Do the NOVs have unfavorable viewing angles?

If the instability strip is really contaminated with non pulsators, the models
for DA white dwarfs and the results derived from these models need to be 
revised. Before taking such a crucial but drastic step, it is important
to make sure the observational results are correct. There are several ways to
falsify a contaminated instability strip. First, the temperatures of the SDSS
sample are derived from spectra with a signal--to--noise rate (SNR) $<$35. 
In this scenario, the temperatures and masses of the non variables could simply
be inaccurate and the stars actually lie outside the instability strip 
(Gianninas, Bergeron, \& Fontaine 2005). There should also be some pulsators
outside the strip due to these uncertainties, and the searches should be 
extended to a wider range of temperatures (e.g. the odd pulsator WDJ 2350-0054
at 10\,350\,K in Mukadam et al. 2004). Second, pulsating white dwarf stars can
appear to be extremely low-amplitude pulsators for a few hours due to the
beating of many excited modes (Kleinman et al. 1998). A third but not the least
important possibility is that variations could have been, in fact, smaller
than the published detection limits, which vary from 2 to 9\,mma (eg. Kanaan et
al. 1992).

The only way to solve the question about the purity of the ZZ Ceti instability 
strip is to observe these stars again through higher SNR spectroscopy and
time-resolved photometry. The higher SNR spectra allow a more accurate 
determination of $T_{\mathrm{eff}}$ and $\log g$ (Kepler et al. 2006). In this
paper, we present our first results on the re-observations of the 
time-resolved photometry of two stars classified as NOVs, finding variability
in their light curves. As every NOV we observed is in fact a low-amplitude
variable, we show strong observational evidence of a pure ZZ Ceti instability
strip.

\section{Observations}

We observed our target list with the SOAR 4.1-m telescope, using the SOAR 
Optical Imager, a mosaic of two EEV 2048$\times$4096 CCDs, thinned and
back-illuminated, with an efficiency around 73\% at 4000\,\AA , at the Nysmith
focus. We observed in fast readout mode with the CCDs binned 4$\times$4, to
decrease the readout+write time to 6.4\,s and still achieve a 0.354"/pixel 
resolution. The exposure times were 30\,s, and we observed up to 4 hours. All
observations were obtained with a Johnson B filter to maximize the amplitude and
minimize the red fringing.

We also observed with the Otto Struve 2.1-m telescope at McDonald Observatory,
using the Argos camera. This camera is mounted at the prime focus of the
telescope, it has a frame transfer CCD, and uses a red cutoff BG40 filter to
reduce the scattered light from the sky (Nather \& Mukadam 2004). The 
integration times were 5\,s and the observations spam for up to 1.4 hours.

Extending our search for new pulsators, we used the 1.6-m telescope at 
Observat\'orio do Pico dos Dias, LNA, in Brazil. For these observations, we
used a frame transfer CCD, with a quantum efficiency of 60\% around 4000\,\AA .
We observed with no filters because the CCD used has no detectable fringing
and the g-mode pulsations in ZZ Ceti stars are coherent at all optical 
wavelengths (Robinson, Kepler, \& Nather 1982). The integration times were
from 40 to 50\,s and spanned up 2.5 hours. In Table~\ref{log1}, we show the
journal of observations. 

For each telescope, we chose the filter set up for which the SNR were
optimal.

\begin{table*}
\caption[]{Journal of observations with 2.1-m telescope at McDonald Observatory,
4.1-m at SOAR, and 1.6-m at OPD.}
\renewcommand{\tabcolsep}{5.2mm}
\begin{tabular}{||c|c|c|c|c|c||}\hline\hline
Star & Run start (UT) & $t_{\mathrm{exp}}$ (s) & $\Delta T$ (hr) & \# points &
Telescope \\ \hline
SDSS J000006.75-004654.0 & 2005-08-04 07:29 & 45 & 2.5 & 199 & 1.6-m\\
 & 2005-08-06 05:23 & 45 & 1.6 & 127 & 1.6-m \\
 & 2005-08-07 05:53 & 50 & 1.2 & 88 & 1.6-m \\
SDSS J030325.22-080834.9 & 2005-12-05 02:59 & 30 & 3.0 & 301 & 4.1-m \\
 & 2005-12-06 02:29 & 30 & 3.5 & 347 & 4.1-m \\
 & 2005-12-07 01:10 & 30 & 4.0 & 400 & 4.1-m \\
SDSS J085325.55+000514.2 & 2005-12-08 06:10 & 30 & 3.3 & 229 & 4.1-m \\
 & 2005-12-09 09:50 & 30 & 3.5 & 170 & 4.1-m \\
SDSS J085507.29+063540.9 & 2005-12-11 05:02 & 30 & 2.8 & 276 & 4.1-m\\
SDSS J091635.07+385546.2 & 2006-01-07 06:35 & 5 & 1.3 & 921 & 2.1-m\\
 & 2006-01-08 08:46 & 5 & 1.4 & 969 & 2.1-m\\
SDSS J165020.53+301021.2 & 2005-08-02 00:37 & 40 & 2.2 & 191 & 1.6-m\\
SDSS J233458.71+010303.1 & 2005-08-06 06:14 & 50 & 2.2 & 306 & 1.6-m \\
 & 2005-08-12 00:36 & 30 & 1.5 & 198 & 4.1-m \\
\hline\hline
\end{tabular}
\begin{list}{Table Notes.}
\item $\Delta T$ is the total length of the each observing run and
$t_{\mathrm{exp}}$ is the integration time of each exposure.
\end{list}
\label{log1}
\end{table*}

The technique used to detect variability is differential time series photometry,
comparing the targets with the other stars in the same field to minimize the 
effects of sky and transparency fluctuations. For each run, we extracted light 
curves using the hsp scripts developed by Antonio Kanaan for IRAF, with weighted
aperture photometry, using different aperture sizes
and selecting the light curve with the highest SNR in the Fourier transform.

For the stars previously classified as NOVs, we had to observe them 
for a minimum of three hours. This minimum time was chosen to avoid 
destructive beating between modes (if there are any that are excited in the 
star) and to be able to reach a lower amplitude limit than previous 
observations. The NOVs we observed had detection limits above 3--4\,mma.

In order to have some objective criterion for determining which peaks are
real in the discrete Fourier transform, we adopted an amplitude limit
such that a peak exceeding this limit only has a 1/1000 probability of being
due to noise (false alarm probability). For these continuous data sets, peaks 
above 3 $\langle A \rangle$ (3 times the square root of the average power) have
a probability smaller than 1 in 1000
of being noise. We checked the detection limit for each star by Monte Carlo
simulation, randomizing the acquisition times.

To know if another peak in the Fourier transform is an intrinsic periodicity of
the star or whether it is only due to the spectral window, we subtracted from 
the light curve, the sine curve with the same amplitude, period, and phase
information as the peak selected
in the Fourier transform. After subtracting it from the light curve, we
recalculated the Fourier transform to verify whether the sine curve was
correct. Then, we repeated the procedure for the remaining periodicities.

\section{New variables}

\begin{figure}
\centering
\includegraphics[angle=0,width=\linewidth]{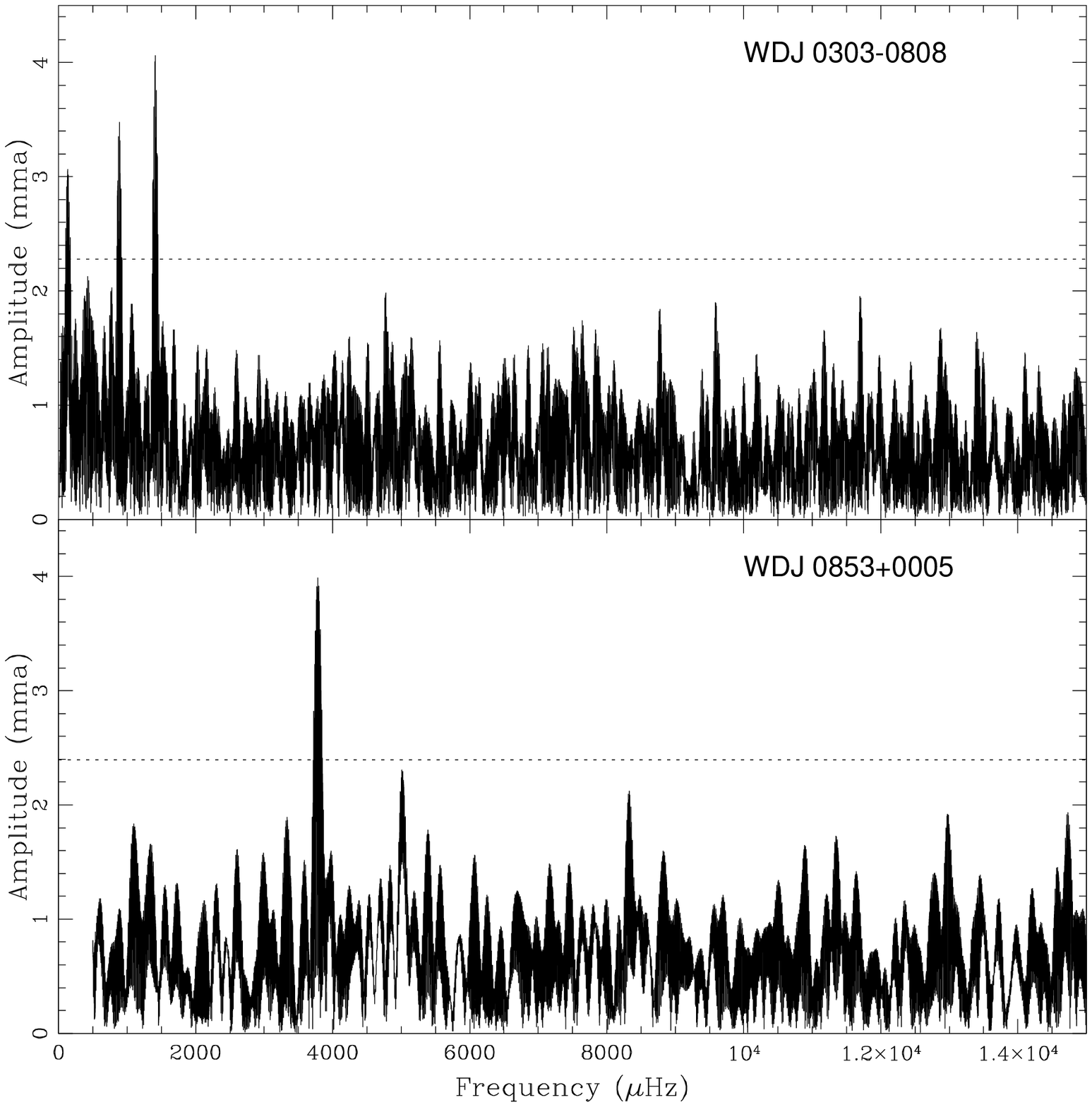}
\caption{Fourier transform of all the data for the stars previously
classified as NOV WDJ 0303-0808 and WDJ 0853+0005.
The dotted lines represent the detection limit above 3 $\langle A \rangle$,
for which peaks have a 1/1000 probability of being due to noise.}
\label{ft2}
\end{figure}

The stars SDSS J030325.22-080834.9 and SDSS J085325.55+000514.2 were listed in
Mukadam et al. (2004) as NOV objects with detection limits of 3--4\,mma. In
Fig.~\ref{ft2}, we show the Fourier transform of the new pulsators previously
classified as constant stars, both being low-amplitude pulsators, with 
amplitudes below the previous detection limit.

\begin{figure}
\centering
\includegraphics[angle=0,width=\linewidth]{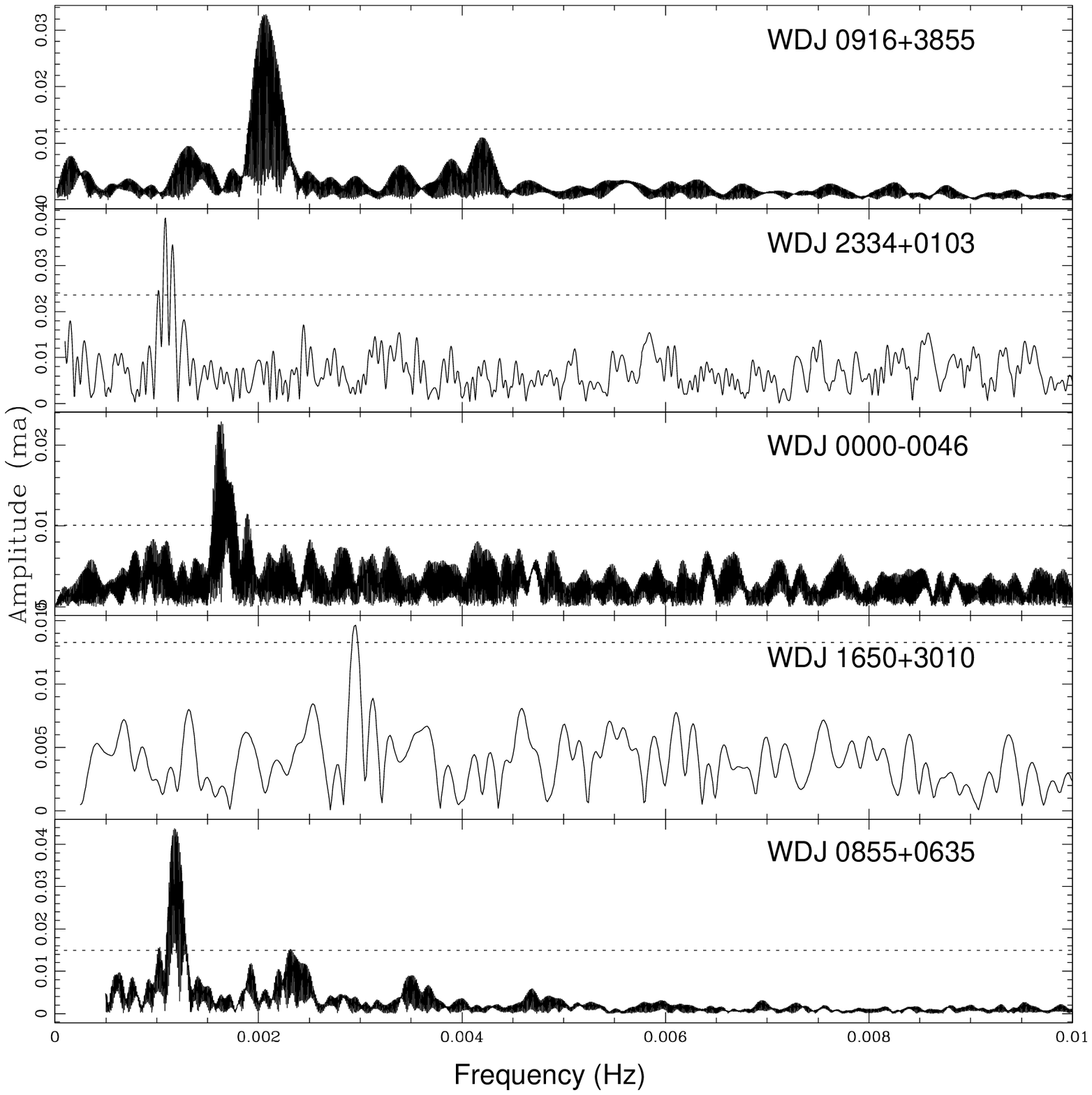}
\caption{Fourier transform of our search for new pulsators using the telescopes
2.1-m at McDonald Observatory, 1.6-m at Observat\'orio do Pico dos Dias, and
the 4.1-m at SOAR. The
candidates were selected by their $T_{\mathrm{eff}}$ derived from comparison
of optical spectra with model atmospheres.
The dotted lines represent the detection limit above 3 $\langle A \rangle$,
for which peaks have a 1/1000 probability of being due to noise.}
\label{tfa}
\end{figure}

We show in Fig.~\ref{tfa} the Fourier transforms of the total light curves of
the new stars discovered to be pulsators that have never been searched before.
These stars were incorporated in our candidate target list based on the same
criterion: their $T_{\mathrm{eff}}$, derived from fitting the SDSS optical 
spectra to Koester's synthetic spectra derived from model atmospheres. We used
a model atmosphere grid similar to the one described by Finley, Koester, \&
Basri (1997), but larger and denser. Within the $T_{\mathrm{eff}}$ range
from 11\,000 to 12\,000~K, we have a 90\% probability of finding pulsators
(e.g. Mukadam et al. 2004). 

Table~\ref{tab2} shows a summary of the physical properties of the new 
pulsators derived by fitting the optical spectra up to 7000\,\AA\, to Koester's
models in the same way as described in Kleinman et al. (2004) and Eisenstein
et al. (2006). We measured the $\chi^2$ on the difference between the observed 
spectra and the models, allowing for a reflux of the models according
to a low-order polynomial. This step was made to correct for effects of
unknown reddening. We used the first seven Chebychev polynomials in a linear
wavelength as our refluxing basis. The uncertainties quoted in the table were
estimated as the internal uncertainty of the fit only, not considering that
$T_{\mathrm{eff}}$ and $\log g$ are correlated. These estimates also do not
consider the changes in $T_{\mathrm{eff}}$ during the pulsation (see Sect.
4.1 for a detailed discussion) or external uncertainties, which are dominant.

\begin{table*}
\caption[]{Physical parameters derived by comparing the whole SDSS optical 
spectra with a grid of model atmospheres and SDSS magnitudes.}
\renewcommand{\tabcolsep}{6.5mm}
\begin{tabular}{||c|c|c|c|c|c|c|c||}\hline\hline
Star & $T_{\mathrm{eff}}$ (K) & $\log g$ & g (mag) & Mass ($M_{\odot}$) &
$T_{\mathrm{eff}}$ (K) ($\langle P\rangle$) \\ \hline
WDJ 0000-0046 & 10\,880$\pm$110 & 8.32$\pm$0.09 & 18.84 & 0.81$\pm$0.06 &
11\,310 \\
WDJ 0303-0808$^1$ & 11\,400$\pm$110 & 8.49$\pm$0.06 & 18.74 & 0.92$\pm$0.04 & 
10\,960 \\
WDJ 0853+0005 & 11\,750$\pm$110 & 8.11$\pm$0.06 & 18.23 & 0.68$\pm$0.04 &
11\,710 \\
WDJ 0855+0635 & 11\,050$\pm$50  & 8.43$\pm$0.03 & 17.25 & 0.88$\pm$0.02 &
11\,140 \\
WDJ 0916+3855 & 11\,410$\pm$50 & 8.10$\pm$0.03 & 16.56 & 0.67$\pm$0.02 &
11\,410 \\
WDJ 1650+3010 & 11\,100$\pm$90 & 8.76$\pm$0.08 & 18.11 & 1.07$\pm$0.04 &
11\,620 \\
WDJ 2334+0103 & 11\,400$\pm$210 & 7.99$\pm$0.14 & 19.24 & 0.60$\pm$0.08 &
10\,930 \\
\hline\hline
\end{tabular}
\begin{list}{Table Notes.}
\item We derived the masses
using $T_{\mathrm{eff}}$ and $\log g$ obtained from optical spectra and
comparing these values with C/O white dwarf evolutionary models (Wood 1995). 
We also calculated the 
$T_{\mathrm{eff}}$ using its relationship with weighted mean pulsation period 
(see Table~\ref{tab3}) derived by Mukadam et al. (2006).
The quoted uncertainties are the internal uncertainties due to fitting
procedures only.
$^1$ Values derived from Gemini spectra are
$T_{\mathrm{eff}}$=11\,960$\pm$160 and $\log g$=8.305$\pm$0.017 using LPT
fit and $T_{\mathrm{eff}}$=11\,420$\pm$110 and $\log g$=7.821$\pm$0.016 when
the fit procedure included the whole spectrum (Kepler et al. 2006).
\end{list}
\label{tab2}
\end{table*}

We analyzed the light curves of each one of the new pulsators to identify their
periodicities, necessary in future seismological analysis for studying their
properties and structure. The list of detected periods in these stars is
shown in Table~\ref{tab3}. 

\begin{table}
\caption[]{Periodicities detected in our discovery runs for the new ZZ Ceti
stars.}
\renewcommand{\tabcolsep}{0.85mm}
\begin{tabular}{||c|c|c||}\hline\hline
Star & Period (s) &  Amplitude (mma) \\ \hline
WDJ 0000-0046 & 611.42; 584.84; 601.35 & 23.00; 15.92; 8.97 \\
WDJ 0303-0808 & 707; 1128 & 4.1; 3.5 \\
WDJ 0853+0005 & 264.35 & 3.99 \\
WDJ 0855+0635 & 850; 433 & 44; 15 \\
WDJ 0916+6108 & 485.09; 447.70; & 32.89; 14.44; \\
& 238.10; 747.20 & 10.76; 9.06 \\
WDJ 1650+3010 & 339.06 & 14.71 \\
WDJ 2334+0103 & 923.15 & 40.37 \\
\hline\hline
\end{tabular}
\label{tab3}
\end{table}

\section{Discussions}

\subsection{Purity of the instability strip}

\begin{figure*}
\centering
\includegraphics[angle=-90,width=\linewidth]{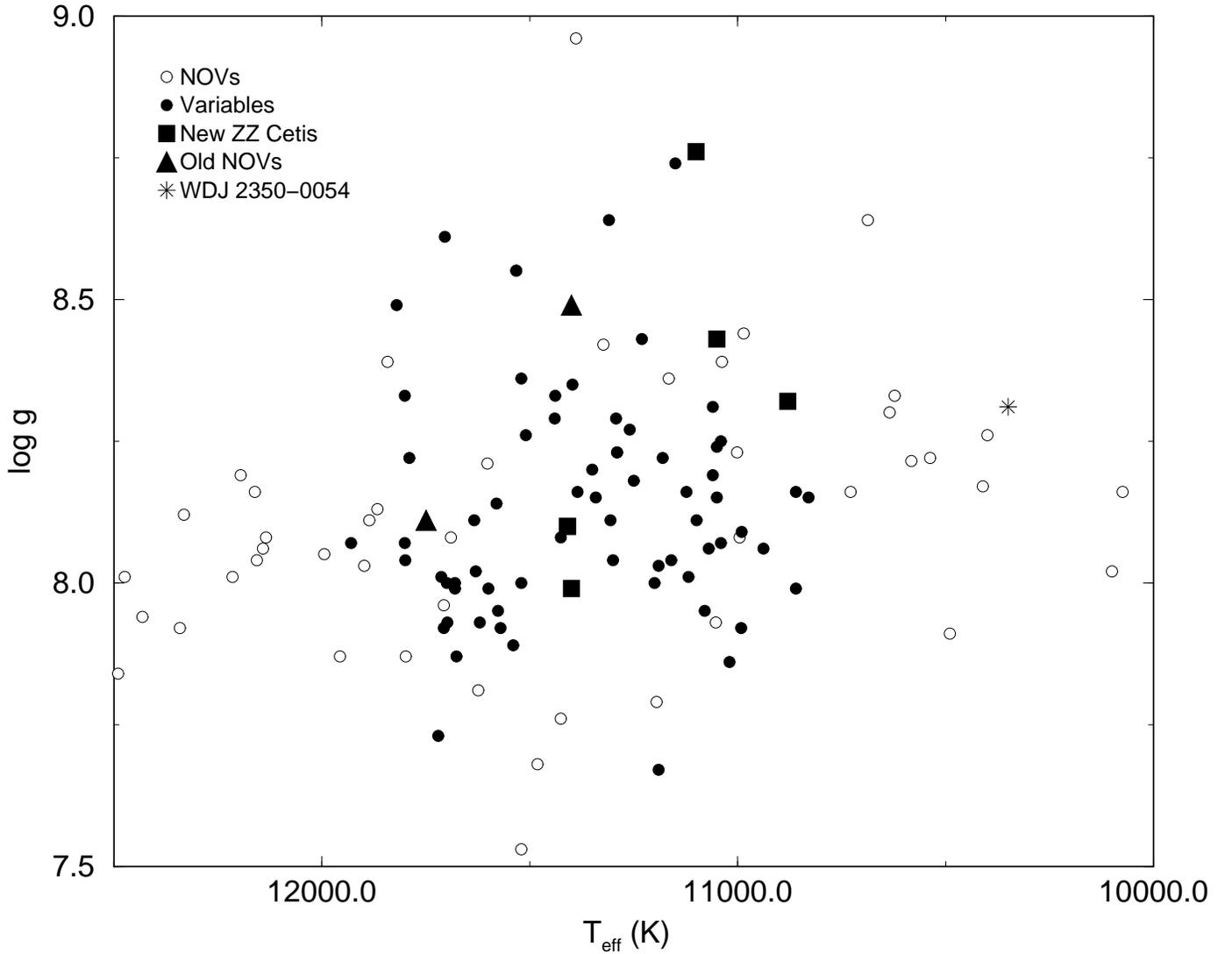}
\caption{Observational instability strip containing only SDSS stars observed for
variability. The filled circles represent the ZZ Cetis and the open circles the
stars for which variation in the light curve was not observed (NOVs) in previous
works that need to be re-observed. The filled squares are the new ZZ Cetis 
reported in this paper, and the filled triangles are the stars that were 
previously classified as NOVs, but we detected variability. The asterisk marks a
variable star that, according to SDSS $T_{\mathrm{eff}}$, is outside the
ZZ Ceti instability strip.} 
\label{inst_old}
\end{figure*}

In Fig.~\ref{inst_old} we show the updated SDSS ZZ Ceti instability strip,
including the NOVs reported by Mukadam et al. (2004) and Mullally et al. (2005)
that still need to be re-observed.  

At this point it is necessary to discuss the star WDJ 2350--0054, which has
a published $T_{\mathrm{eff}}=10\,350\pm60$ (Mukadam et al. 2004), indicated
by an aster-ix in Fig.~\ref{inst_old}. It is
a pulsating star outside the instability strip with
$T_{\mathrm{eff}}$ significantly cooler than the red edge. 
Because there is no 
reasonable explanation for such a low temperature, we will not consider this
star as the red edge of the instability strip. There are three independent
SDSS spectra for this star, all of them providing a lower $T_{\mathrm{eff}}$
than does the red edge when the whole spectrum is used in the fitting procedure.
However, our line profile technique (LPT) fit, which has a larger 
scatter in the $T_{\mathrm{eff}}$
determination for the SDSS spectra because of the lower SNR in the lines, the
resulting $T_{\mathrm{eff}}$ brings this particular star inside the
instability strip. High SNR spectra are necessary to reliably
determine its physical parameters.

We detected variability in two stars previously classified as NOVs. There are 
still 19 NOVs inside the instability strip, which need to be observed again. A
strong argument for a pure instability strip is that
every star we re-observed for variability turned out to be a pulsator.

Kepler et al. (2005) discuss the external uncertainty measured from SDSS
duplicate spectra, showing the uncertainty is around 300\,K in 
$T_{\mathrm{eff}}$ and 0.21 dex in $\log g$, a significant portion of the size 
in $T_{\mathrm{eff}}$ of the instability strip ($\sim 1\,400\,K$).
There is also another important effect on the contamination of a pure 
instability strip with non-pulsators: temperature variation during a pulsation
cycle. For a low-amplitude low $\ell$ pulsator, 
the changes should be rather small, 
around 50\,K, but for larger amplitude pulsators, temperature variation can
reach 500\,K (Robinson, Kepler, \& Nather 1982). 

The Gianninas, Bergeron, \& Fontaine (2005) simulation of an LPT fitting shows 
that uncertainties for the SDSS spectra are on the order of 450\,K for the
average SNR$\sim$20 in our sample. These are external uncertainties of the
LPT method; however, we used the whole spectra fitting.

Kepler et al. (2006) report results from Gemini spectra with 
SNR around 100 for four stars from the SDSS sample. They show that their new
$T_{\mathrm{eff}}$ were similar to the ones obtained with SDSS spectra. 
The SDSS quoted uncertainties are underestimated by 60\% in $T_{\mathrm{eff}}$
and by a factor of 4 in $\log g$, and the correlation between $T_{\mathrm{eff}}$
and $\log g$ is an important factor. Their simulation indicates that fitting 
using the whole spectra with reliable flux calibration is better than only an
LPT fit.

All these uncertainties added together can falsify our interpretations
of the impurity of the instability strip. Higher SNR spectra averaging
out pulsations (e.g. Bergeron et al. 2004), as well as longer exposure 
time-series imaging, are required for lowering the uncertainties in temperature 
and the detection limits of pulsations for all stars within the boundaries of 
the ZZ Ceti instability strip.

\subsection{Changes in pulsation spectrum with temperature}

While the star cools down inside the instability strip, its pulsation
spectrum changes, as do the pulsation amplitudes. Evolving from the blue
(hot) to the red (cold) edge, the periods get longer and their
amplitudes larger. For example, the hottest known pulsator, 
G~226--29 (Gianninas, Bergeron, \& Fontaine 2005), has a dominant
mode at 109\,s and an amplitude of around 3\,mma (Kepler et al. 1995) in the 
optical and no other periodicities besides the components of this triplet. On
the other hand, the cooler star G~29--38 has 20 identified modes (Kleinman et
al. 1998), and its pulsation spectra change dramatically from run to run. In one
season, this star showed no detectable pulsation; but its dominant mode is
normally around
700\,s, and the pulsation amplitude for a single mode can reach 60\,mma. All
five of the new variables presented in Fig.~\ref{tfa} are long-period 
high-amplitude pulsators, while the previous NOVs have low amplitude, 
independent
of their periods.

The qualitative explanation for the change in pulsation spectra within the
ZZ Ceti instability strip is based on the behavior of the convection zone,
starting at the base of the partial ionization layer, which moves
inwards as the star cools. The convection zone gets bigger, allowing more
modes to be excited, as the star progresses to the red edge. Because the
base of the convection zone is deeper, the thermal time is longer, so
the excited periodicities are rather long. In addition, as the position of 
the convection zone is more internal for cooler stars, more energy is
needed to move the upper layers, which leads to higher amplitudes. 
Note that the timescale for crossing the instability strip is almost 1\,Gyr.

Even though we understand how pulsations start, because of the increase in 
opacity 
at the developing partial ionization zone, we do not understand how the
pulsations stop. All evolutionary models predict a much cooler red edge than
the coolest ZZ Ceti observed, even though Wu \& Goldreich (1999) propose that
pulsation should stop for periods around 1\,400\,s (as observed) in their
models of convection-dominated DA pulsations. Kanaan, Kepler, \& Winget (2002) 
looked for low-amplitude
pulsators at the red edge, which would represent when the ZZ Cetis stop
to pulsate, but they found none. Mukadam et al. (2006) found seven low amplitude
pulsators with long periods, which is too large a number to be explained by 
inclination
effects, especially when we consider that different $m$'s cancel at opposite 
inclination angles. In our sample, WDJ 0303-0808, shows long periods
and low amplitude.

Another reason for low amplitudes occurs for high mass stars, when their
interiors are partially crystallized, like for BPM 37093 (Kanaan, Kepler, \& 
Winget 2005), as the pulsations are restricted to the upper layers above 
the crystallized core. However, to test this hypothesis for the low amplitude
long period star observed, it is necessary to 
obtain spectra with higher SNR than the SDSS spectra, specially in the
region of the lines H8 and H9, because they are more sensitive to gravity,
so measuring $T_{\mathrm{eff}}$ and $\log g$ accurately.

Mukadam et al. (2006), following Kanaan, Kepler, \& Winget (2002), discuss the
relationship between weighted mean period and
spectroscopic temperature, after studying the changes in pulsation properties
across the ZZ Ceti instability strip. They found a linear relation between
these two physical quantities for all ZZ Ceti stars with available temperatures.

Among the new pulsators, the two stars classified as NOVs are low
amplitude pulsators, but they have completely different main periodicities. 
WDJ 0303-0808's detected periodicities are 707 and 1128\,s, typical of
the red edge cool pulsator, while 
WDJ 0853+0005's main mode is around 264\,s, typical of a blue edge pulsator. 
Either 
WDJ 0303-0808 is a high mass star, where crystallization already occurs
in its interior or it is an example of a pulsator stopping to pulsate. 
In Table~\ref{tab2}, we listed $T_{\mathrm{eff}}$ and $\log g$ derived from
SDSS spectrum. Using these values and the radius calculated by 
Althaus et al. (2005), we derived a stellar mass of 0.93\,$M_{\odot}$. Kepler
et al. (2006) derived a lower stellar mass of 0.83\,$M_{\odot}$ and 
0.52\,$M_{\odot}$ from Gemini spectra fitting with LPT and whole spectra,
respectively.
As crystallization is important within the ZZ Ceti instability strip only for 
stars with masses higher than $\sim 1\,M_{\odot}$, when a significant portion
of the star will be crystallized, this star is most likely an example of a star
stopping to pulsate. On the other hand, the high $\log g$ determined from SDSS
spectra would indicate that the star WDJ 1650+3010 could be crystallized; 
however, its observed amplitude does not indicate that. 
The known correlation between $T_{\mathrm{eff}}$ and $\log g$ determination
could again explain this issue, as the determinations were based on low
SNR spectra.

Using the Mukadam et al. (2006) relation, the $T_{\mathrm{eff}}$ for the star
WDJ 0853+0005 should be 11,700\,K in agreement with the one derived from the
low SNR spectra 11,700\,K. For WDJ 0303-0808, the pulsation periods indicate
11,000\,K, compared with 11,400\,K from the spectra. 
More observations
of the star WDJ 0303-0808 are required, both with high SNR spectroscopy and 
time-resolved photometry, for a better understanding of its
structure.

\section{Concluding remarks}

We report the discovery of seven new pulsating DA stars,
bringing the total to 126 known ZZ Cetis in the narrow
temperature range 12\,270 K $\geq T_{\mathrm{eff}} \geq$ 10\,850 K,
corresponding to the existence of a
partial ionization of hydrogen and development of
a sub-surface convection zone. 
As two of these stars were previously reported as NOV, we need to
lower the detection limits for all other NOVs observed by 
Mukadam et al. (2004) and Mullally et al. (2005), map the instability strip 
in both 
$T_{\mathrm{eff}}$ and $\log g$ using accurate determinations of these
parameters for the new variables and NOVs, for which we need higher SNR 
than those
achieved in the SDSS spectra.

\begin{acknowledgements}
The authors acknowledge the support of CNPq fellowships and a NASA origins 
grant.
\end{acknowledgements}

%
%

\end{document}